\documentclass[twocolumn, nofootinbib, showpacs]{revtex4-1}

\usepackage{amssymb}
\usepackage{amsmath}
\usepackage{amsfonts}
\usepackage{dsfont}
\usepackage[usenames,dvipsnames]{xcolor}
\usepackage[pdftex]{graphicx}
\usepackage{bm}
\usepackage[pdftex]{hyperref}
\hypersetup{plainpages=false,colorlinks=true,linkcolor=blue, citecolor=blue, urlcolor=blue}

\usepackage[english]{babel}
\usepackage{xcolor}
\usepackage{bbm}
\usepackage{bbold}
\usepackage{mathrsfs}
\usepackage{natbib}

\begin{document}

\title{Hall effect in cuprates with incommensurate collinear spin-density wave}
\author{M. Charlebois$^{1}$, S. Verret$^{1}$, A. Foley$^{1}$, O. Simard$^{1}$, D. S\'en\'echal$^{1}$, A.-M. S. Tremblay$^{1,2}$}
\affiliation{$^1$ D\'epartement de Physique and Institut quantique, Universit\'e de Sherbrooke, Sherbrooke,
QC, Canada\\
$^{2}$Canadian Institute for Advanced Research, Toronto, Ontario, Canada.}
\date{\today}

\begin{abstract}
The presence of incommensurate spiral spin-density waves (SDW) has been proposed to explain the $p$ (hole doping) to $1+p$ jump measured in the Hall number $n_H$ at a doping $p^*$. Here we explore incommensurate {\it collinear} SDW as another possible explanation of this phenomenon, distinct from the incommensurate {\it spiral} SDW proposal. We examine the effect of different SDW strengths and wavevectors and we find that the $n_H\sim p$ behavior is hardly reproduced at low doping. Furthermore, the calculated $n_H$ and Fermi surfaces give characteristic features that should be observed, thus the lack of these features in experiment suggests that the incommensurate collinear SDW is unlikely to be a good candidate to explain the $n_H\sim p$ observed in the pseudogap regime.
\end{abstract}

\maketitle

%\hyphenation{Brillouin}

%\pacs{72.15.Jf, 71.27.+a, 73.43.Cd}

\section{Introduction}

Recently, a measurement of the Hall effect in YbBa$_{2}$Cu$_{3}$O$_{y}$ (YBCO) by Badoux et al.~\cite{badoux_change_2016} provided some clues on the zero-temperature normal state that is found when a magnetic field prohibits superconductivity. A sharp jump in the effective carrier density from $p$ (hole doping) to $1+p$ was observed around $p^*\sim 0.19$, the extrapolated zero-temperature value of the pseudogap line  $T^*(p)$.~\cite{tallon_doping_2001} % along with a Fermi surface reconstruction. 
It was suggested that this is an important clue to understand the pseudogap phenomenon.

Since then, an appreciable number of phenomenological theories were proposed to explain this behavior. Most of them %being able to 
reproduced the jump in the effective carrier number measured from the Hall effect $n_H$. The candidate theories can be separated in two groups: those based on a hypothetical long range magnetic order and those based on Mott-like physics.

In the first group, a simple antiferromagnet~\cite{storey_hall_2016} was shown %to be
sufficient to reproduce the $p$ behavior at low doping. However, in experiments, antiferromagnetism does not extend above $p=0.05$~\cite{haug_neutron_2010} and therefore this scenario is unlikely. % no antiferromagnetism should survive, which is a strong argument against this approach.
Spiral antiferromagnets, commensurate and incommensurate, were also studied~\cite{eberlein_fermi_2016-1}. In the incommensurate case (above $p=0.05$), hole pockets twice as large as in the simple antiferromagnet were predicted. This would show up in quantum oscillations.

In the second group of theories, based on Mott physics, the resonating-valence-bond spin-liquid ansatz of Yang, Rice and Zhang (YRZ)~\cite{yang_phenomenological_2006}, a phenomenological model of the pseudogap, was able to reproduce the jump in Hall carrier~\cite{storey_hall_2016}. An implementation of the fractionalized Fermi liquid theory (FL$^*$) was also able to reproduce this jump~\cite{chatterjee_fractionalized_2016}. 

%(synonym, ouin, il faut peut etre que je change la formulation).

All of the above theories can be expressed as two-band effective models~\cite{verret_phenomenological_2017}. In other words, a strong enough order, introduced as an effective mean field, opens a gap at half-filling, splitting a single band into two bands. This regime is associated with an effective carrier density $p$. If this mean field order is removed, one recovers the original single band with an effective carrier density of $1+p$. Each theory that reproduces the Hall jump~\cite{badoux_change_2016} tunes this mean field as a function of doping in order to recover the two bands ($p$) at low doping and a single band ($1+p$) above $p^*$. The SU(2) theory gave a different explanation of this same jump.~\cite{morice_evolution_2017} 
%If we tune this field as a function of doping. Around half-filling, when the field is present, we get two bands, hence a $p$ behavior and at high doping, when the field vanishes, we get only one band, hence a $1+p$ behavior. 
%They all reproduce qualitatively the result from Badoux et al.~\cite{badoux_change_2016}. 

In this paper, %we don't want to discriminate which of these models is the correct one, but 
we %want to 
explore another possibility: the incommensurate collinear spin-density wave (SDW, see Fig.~\ref{figureSDW})~\cite{christensen_nature_2007}. By collinear, we mean a SDW which is a modulation of the amplitude of the spin order parameter by contrast with the spiral SDW, which is a rotation of the spin with constant amplitude~\cite{tung_ab_2011,tsunoda_spin-density_1989}.  It is not clear experimentally whether the SDW is spiral or collinear, as discussed in the context of  La$_{2-x}$Sr$_x$CuO$_4$ (LSCO) measurements~\cite{christensen_nature_2007}. But we do know that for $p>0.05$, there is an incommensurate SDW that survives at low temperature, either collinear or spiral. This has been found by neutron scattering in LSCO % La$_{2-x}$Sr$_x$CuO$_4$ (LSCO) 
and %YbBa$_{2}$Cu$_{3}$O$_{y}$ (YBCO)
YBCO.\cite{yamada_doping_1998,haug_neutron_2010,fujita_static_2002}

Since calculations for the spiral SDW have already been done~\cite{eberlein_fermi_2016-1}, we focus only on the long-range incommensurate collinear SDW. %on the Hall number $n_H$ as a function of $p$ in YBCO. 
The tight-binding Hamiltonian along with the formalism used to evaluate the Hall number $n_H$ is shown in section~\ref{section:model}. 
In section ~\ref{section:results}, we present results following a very gradual approach: we compute $n_H$ as a function of $p$, % for different cases. We begin with no
first without SDW, then with a commensurate SDW, and finally %replace it 
with %an 
incommensurate SDW. 
This progression reveals
%We proceed this way in order to see progressively 
the effect of each modification and builds a general understanding that will be useful for our discussion (section~\ref{discussion}). In the end we show how unlikely 
it is that the incommensurate collinear SDW explains the jump in Hall carrier. We conclude that if an order is associated with the pseudogap at $T=0$, %it (is probably ...)
%We conclude that the pseudogap order, associated to the $p$ behavior, 
it is probably best represented by a two-band effective model.

%One of the key elements we will look at throughout the whole paper is the $p$ 

\section{Model}
\label{section:model}
%\subsection{Commensurate}

We use the following tight-binding Hamiltonian: 
\begin{align}
H=&
\sum_{\mathbf{k},\sigma}
\xi_\mathbf{k} c^{\dag}_{\mathbf{k},\sigma} c^{\vphantom{\dag}}_{\mathbf{k},\sigma}
+
M
\sum_{\mathbf{k},\sigma} \sigma
(c^{\dag}_{\mathbf{k},\sigma} c^{\vphantom{\dag}}_{\mathbf{k}+\mathbf{Q},\sigma} +\text{H.c.}).
\label{Hsdw}
\end{align}
The first term is the kinetic energy, where the dispersion relation $\xi_{\mathbf{k}}$ is defined with first-, second- and third-neighbor hopping energy $t$, $t^\prime$ and $t^{\prime\prime}$:
\begin{align}
%\xi_{\mathbf{k}} = & \;\epsilon_{\mathbf{k}} -\mu \\
\xi_{\mathbf{k}} = 
& -2t (\cos (k_x) + \cos (k_y)) \nonumber \\
& -2t^{\prime} (\cos (k_x + k_y) + \cos (k_x - k_y)) \nonumber \\
& -2t^{\prime\prime} (\cos (2 k_x) + \cos (2 k_y)) - \mu.
\label{basicDispersion0}
\end{align}
The second term of Hamiltonian~\eqref{Hsdw} is the SDW mean-field energy with amplitude $M$. $c^{\dag}_{\mathbf{k},\sigma}$ and $c^{\vphantom \dag}_{\mathbf{k},\sigma}$ are the creation and annihilation operators of momentum $\mathbf{k}$. $\mathbf{Q}$ is the wave vector of the SDW.\footnote{Only considering $\mathbf Q$ is sufficient to generate smaller gaps at the harmonics $2 \mathbf Q$, $3 \mathbf Q$ and so on, through the diagonalization of the Hamiltonian.}
$\sigma=\pm 1$ is the spin index. We work in units where Planck's constant and lattice spacing are unity.
% One just need to keep in mind that the SDW amplitude of spin up must be the opposite of the one in spin down.

\begin{figure}[t]
\includegraphics[width=1.00\columnwidth]{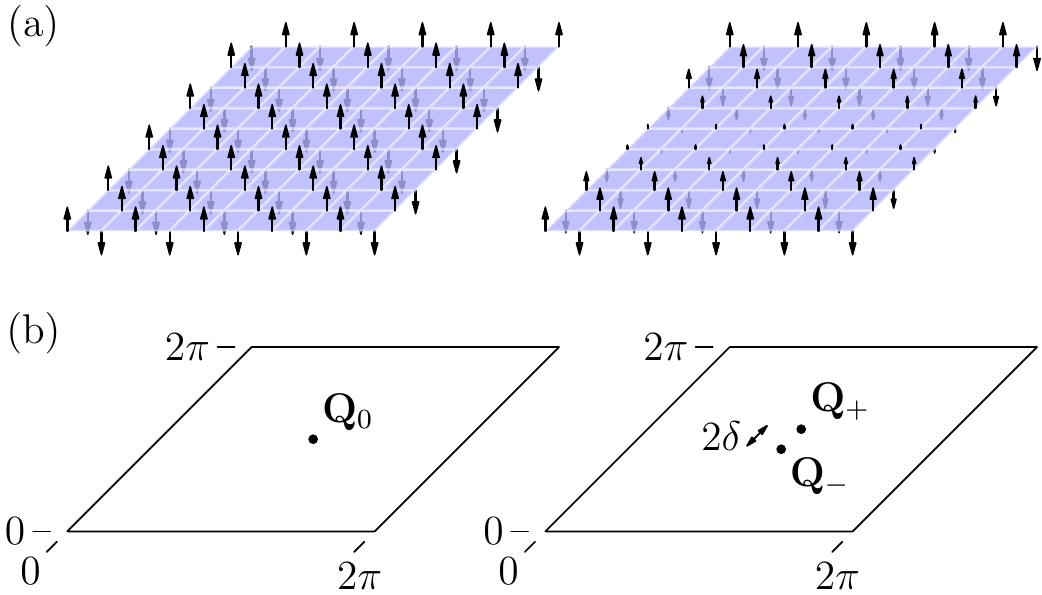}
\caption{(a) Local moment distributions for commensurate $(\pi,\pi)$ collinear SDW (left) and incommensurate collinear SDW (right). The incommensurate SDW is the equivalent to a commensurate SDW modulated by $\cos(2\pi\delta \mathbf{\hat{y}})$. Here, $L=9$ and $\delta=\tfrac{1}{18}$ (see Sec.~\ref{sectionB} for definitions). (b) $\mathbf{Q}$ in the reciprocal space for commensurate collinear SDW (left) and incommensurate collinear SDW (right).} 
\label{figureSDW}
\end{figure}

The commensurate $(\pi,\pi)$ SDW is presented in section~\ref{sectionA} and the incommensurate collinear SDW in section~\ref{sectionB}. Both SDW are shown on Fig.~\ref{figureSDW}. It is important to emphasize that we do not solve the truly incommensurate case but only rational approximations, namely  commensurate SDW with shorter or longer periods, depending on the definition of~$\mathbf Q$. This distinction between commensurate $(\pi,\pi)$ SDW and incommensurate SDW is consistent with common usage in experiments.

%It is important ot emphasize that the only Hamiltonian~\eqref{Hsdw} we can solve using our model yields commensurate SDW with shorter or longer periods, depending on the definition of~$\mathbf Q$. 

\subsection{Commensurate case}
\label{sectionA}
When the wave vector is
\begin{align}
\mathbf Q_0 = \big( \pi, \pi \big),
\label{Q0}
\end{align}
%we say that 
the SDW is commensurate with period $2$ in $x$ and $y$ directions; it corresponds exactly to the spin ordering of an N\'eel antiferromagnet. In that case, we can define the following two-orbital spinor
\begin{align}
\Psi^{\dagger}_{\mathbf k,\sigma} = 
\begin{pmatrix}
c^{\dagger}_{\mathbf k, \sigma} &
c^{\dagger}_{\mathbf k+\mathbf Q_0, \sigma}
\end{pmatrix},
\end{align}
and matrix Hamiltonian
\begin{align}
\hat{H}_{\mathbf k, \sigma}=
\begin{pmatrix} 
  \xi_\mathbf k & \sigma M  \\
  \sigma M & \xi_{\mathbf k+\mathbf Q_0}
\end{pmatrix},
\end{align}
so that the original Hamiltonian~\eqref{Hsdw} can be expressed as: %in this new basis:
\begin{align}
H=&
\sum_{\substack{\mathbf k  \in \text{rBz}\\ \sigma}}
\Psi^{\dagger}_{\mathbf k,\sigma} \hat{H}_{\mathbf k,\sigma} \Psi^{\vphantom \dagger}_{\mathbf k,\sigma}.
\label{Hsdw2}
\end{align}
%\begin{align}
%H=&
%\sum_{\mathbf k}^{\text{rBz}}
%\sum_{\sigma}
%\Psi^{\dagger}_{\mathbf k,\sigma} \hat{H}_{\mathbf k,\sigma} \Psi^{\vphantom \dagger}_{\mathbf k,\sigma}.
%\label{Hsdw2}1
%\end{align}
The sum is restricted to the reduced Brillouin zone (rBz) to avoid double counting. In the commensurate case, this rBz corresponds to the antiferromagnetic Brillouin zone. The eigenenergies $E_{\mathbf k,n}$ (for band $n$) are simply obtained through diagonalization of the $H_{\mathbf k, \sigma}$ matrix. 

\subsection{Incommensurate case}
\label{sectionB}
In cuprates, the SDW does not remain commensurate for every doping. Beyond a threshold, it becomes incommensurate. The single SDW vector $\mathbf Q_0$ then splits locally in two wave vectors:
\begin{align}
\mathbf Q_\pm = 2\pi \Big( \frac{1}{2}, \frac{1}{2} \pm \delta \Big)
\label{unidirectionalQ}
\end{align}
as experimentally measured with neutron scattering on single crystals~\cite{haug_neutron_2010,fujita_static_2002} (see Fig.~\ref{figureSDW}). Higher order harmonics are negligible. Note that this order breaks $C_4$ rotational symmetry. %Both these measurements and the Hall coefficient measurements~\cite{badoux_change_2016,segawa_intrinsic_2004,ando_evolution_2004} were made on single crystals, thus there is no domain averaging expected 
%thus our choice of anisotropic $\mathbf Q_\pm$ is justified.

%Note that most measurements report a splitting in four directions, but since the probes are sensitive to the bulk, it is assumed that the splitting is locally unidirectional (?? citation). 

We can generalize the approach above in a straightforward way for incommensurate SDW by defining the spinor % of dimension $L_x \times L_y$ (the period of the system in both direction $x$ and $y$ respectively):
of dimension $2L$:
\begin{align}
\Psi^{\dagger}_{\mathbf k,\sigma} = 
\begin{pmatrix}
c^{\dagger}_{\mathbf k,\sigma} & ... &
c^{\dagger}_{\mathbf k+m\mathbf Q_+,\sigma} & 
... & 
\end{pmatrix},
\end{align}
where $m$ ranges from $0$ to $2L-1$. $2L$ is an even integer that defines the denominator of the fraction of the incommensurability:
\begin{align}
\delta = \frac{q}{2L},
\end{align}
with $q$ an integer. With this spinor definition, the original Brillouin zone is then separated in $2L$ rBz.
Hence, with $2L$ additions of the vector $\mathbf Q_+$, modulo a vector of the reciprocal lattice,\footnote{Of the original Brillouin zone.} we cycle through every different rBz. Note that we need an even number of additions of $\mathbf Q_+$ in order to cycle through every different rBz. %The shape of every rBz is a rectangle of diagonal $\mathbf Q_+$. 
We could use $\mathbf Q_-$ and it would cover the exact same $2L$ rBz since $\mathbf Q_+=-\mathbf Q_-$ modulo a vector of the reciprocal lattice. %However, one need to use only $\mathbf Q_+$ or $\mathbf Q_-$, but not both, to cover every rBz.

The Hamiltonian matrix $\hat{H}_{\mathbf k, \sigma}$ in this basis is of dimension $2L$ by $2L$. It has $\epsilon_{\mathbf k +m\mathbf Q_+}$ on the diagonal and zero on most of the off-diagonal elements. When the column index $m$ and the row index $m^\prime$ are such that $m-m^\prime$ modulo $2L$ is $\pm 1$, the matrix element is the scalar $\sigma M$. This matrix is almost, but not quite, tridiagonal due to the finite term at indices $(m,m^\prime)=(1,2L)$ and $(2L,1)$.

We name this representation where $\delta\ne 0$ ``incommensurate SDW''. However, in reality it is commensurate with a long period. In other words, since $\delta$ is a fraction, the only order that can be represented by our model repeats every $2L$ sites in the $y$ direction and every $2$ sites in the $x$ direction. %The larger $L_y$ is, the more lattice sites it takes to repeat in $y$, but it always repeats. Note that the order is long range since it repeats periodically. 

%Note that, as defined here (repeat periodically, no real-space damping), the order is long range

%Note that the order is long-range since it repeats periodically (without any real space damping). 

%Since we don't add any envelope, the order repeats itself so it is long ranged.

%

%in the sense that the SDW repeats itself every two lattice site in the $x$ direction of the crystal and every $L$ lattice site in the $y$ direction. 
%Our model is not purely incommensurate in the sense that it represents a long range order SDW. The larger $L_y$ is, the more lattice site it takes to get the same pattern in $y$, but it will repeat due to the implicit periodic boundary condition. 

Note that with $\delta=0$, we recover the commensurate model of the previous section. 

\subsection{Formula for the Hall conductivity}

With $\hbar=1$, the electron charge $−e$ and the normalization volume
$V$, the Hall number $n_H$ and resistivity $R_H$ are ~\cite{voruganti_conductivity_1992,verret_phenomenological_2017}:
\begin{align} 
R_H = \frac{\sigma_{xy}}{\sigma_{xx}\sigma_{yy}} = \frac{V}{e n_H},
\end{align}
where $\sigma_{xx}$ is the longitudinal conductivity at zero temperature in the zero-frequency limit when interband transitions can be neglected
\begin{align} 
\sigma_{xx}=\frac{ e^2 \pi  }{V}\sum_{\substack{n \\ \mathbf k  \in \text{rBz}}}
\Big(
\frac{\partial E_{\mathbf{k},n}}{\partial k_x}\Big)^2
A^2_{\mathbf k,n}(0),
\end{align}
and $\sigma_{xy}$ is the transversal conductivity~\cite{voruganti_conductivity_1992,storey_hall_2016,verret_phenomenological_2017,eberlein_fermi_2016-1}
:
\begin{align} 
\sigma_{xy}=\frac{e^3 \pi^2}{3V}\sum_{
%\mathbf k  \in \text{rBz},n} 
\substack{n \\ \mathbf k  \in \text{rBz}}
}
\Big[
-&2 
\frac{\partial E_{\mathbf{k},n}}{\partial k_x}
\frac{\partial E_{\mathbf{k},n}}{\partial k_x}
\frac{\partial^2 E_{\mathbf{k},n}}{\partial k_x \partial k_y}
\\
+\Big(
\frac{\partial E_{\mathbf{k},n}}{\partial k_x}\Big)^2
\frac{\partial^2 E_{\mathbf{k},n}}{\partial k_y^2}+ 
&\nonumber
\Big(
\frac{\partial E_{\mathbf{k},n}}{\partial k_y}\Big)^2
\frac{\partial^2 E_{\mathbf{k},n}}{\partial k_x^2}
\Big] A^3_{\mathbf k,n}(0).
\end{align}
$E_{\mathbf{k},n}$ is the eigenenergy of band $n$. Here, the band index $n$ includes the spin index $\sigma$. $A_{\mathbf k,n}(\omega)$ is the spectral weight for band $n$:
\begin{align}
A_{\mathbf k,n}(\omega)&\equiv
-\frac{1}{\pi} \text{Im}\Bigg(\frac{1}{\omega + \text{i}\eta - E_{\mathbf k,n}} \Bigg)
.
\label{eq_akw}
\end{align}
The Lorentzian broadening $\eta$ is necessary for the integral to converge and corresponds to constant lifetime $\tau=\tfrac{1}{2\eta}$. We choose $\eta=0.05$. However, a different value with the same magnitude yields similar results~\cite{verret_phenomenological_2017}.

The derivatives
$\tfrac{\partial E_{\mathbf{k},n}}{\partial k_\alpha}$
are the Fermi velocities in the $\alpha=x,y$ direction and 
$\tfrac{\partial^2 E_{\mathbf{k},n}}{\partial k_\alpha \partial k_\beta}$
corresponds to the $\alpha\beta$ component of the inverse effective mass tensor. When the above formulae are used, it is important to use the derivatives of the eigenenergies $E_{\mathbf{k},n}$ and not of the bare band $\xi_{\mathbf{k}}$ to capture the correct behavior of the Hall effect~\cite{voruganti_conductivity_1992,storey_hall_2016,eberlein_fermi_2016-1,verret_phenomenological_2017}. For an arbitrary given $\mathbf k$ point, it is easy to compute the $\hat{H}_{\mathbf k,\sigma}$ matrix and find its eigenenergies $E_{\mathbf{k},n}$. However, it is more complicated to obtain their derivatives
%the derivatives are more complicated to obtain 
%if we deal with an incommensurate SDW.
for matrices larger than 
$2\times2$.
Appendix~\ref{derivativesSection} shows a systematic approach to calculate exactly these derivatives with a single diagonalization of the matrix $\hat{H}_{\mathbf k,\sigma}$ at each $\mathbf k$ point by a generalization of the Hellmann-Feynman theorem~\cite{feynman_forces_1939,deb_note_1972}. 

% Note that we could always estimate with finite differences, but there is a possible imprecision around degeneracies due to the arbitrary ordering of the eigenenergies. 

%\begin{align} 
%\sigma_{xy}=\frac{\pi^2 B^z}{2N}\sum_{\mathbf k,n} 
%\Big[
%&
%\big(
%v^n_{x}(\mathbf k)\big)^2
%v^n_{yy}(\mathbf k)+ 
%\nonumber
%\big(
%v^n_{y}(\mathbf k)\big)^2
%v^n_{xx}(\mathbf k)
%\\
%-2& 
%v^n_{x}(\mathbf k)
%v^n_{y}(\mathbf k)
%v^n_{xy}(\mathbf k)
%\Big] A_{\mathbf k,n}(0)
%\end{align}
%
%
%\begin{align} 
%\sigma_{xy}=\frac{\pi^2 B^z}{2N}\sum_{\mathbf k,\alpha} 
%\Big[
%&
%(v_{x}^\alpha(\mathbf k))^2
%\frac{\partial v_{y}^\alpha(\mathbf k)}{\partial k_y}+ 
%\nonumber
%(v_{y}^\alpha(\mathbf k))^2
%\frac{\partial v_{x}^\alpha(\mathbf k)}{\partial k_x}
%\\
%-2&
%v_{x}^\alpha(\mathbf k)
%v_{y}^\alpha(\mathbf k)
%\frac{\partial v_{y}^\alpha(\mathbf k)}{\partial k_x} 
%\Big] ( A^\alpha_{\mathbf k}(0) )
%\end{align}

\section{Results}
\label{section:results}
In this section, we look at the Hall number $n_H$ as a function of hole doping $p$ relative to half-filling. We choose this convention to match the experimental data and theoretical studies in hole doped cuprates~\cite{badoux_change_2016,balakirev_quantum_2009,storey_hall_2016,eberlein_fermi_2016-1}. With this convention, the $p$ axis goes from a completely filled band at $p=-1$ to an empty band at $p=1$, with $p=0$ corresponding to half-filling. We will study in detail three cases in order to understand progressively the complications of the underlying physics and to familiarize ourselves with the general behavior of $n_H$ versus $p$ curves. 

\subsection{No SDW}

\begin{figure}[t]
\includegraphics[width=1.00\columnwidth]{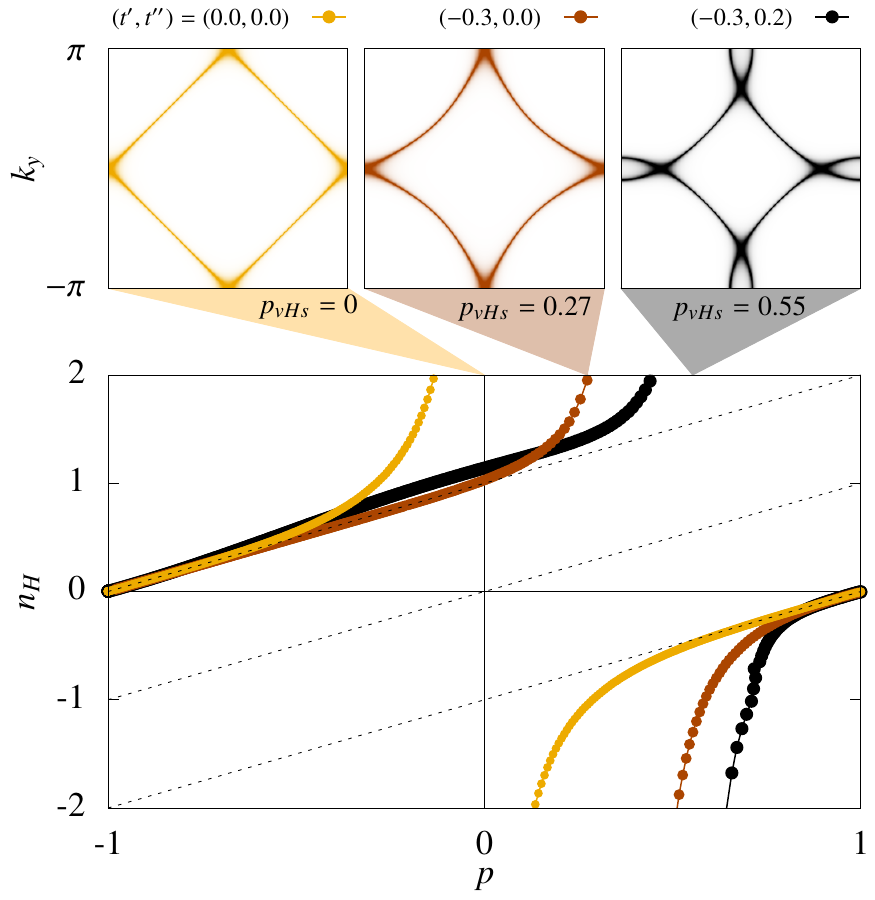}
\caption{(bottom) Hall conductivity $n_H$ as a function hole doping $p$ for $3$ different band parameters. (top) Fermi surfaces ($\sum_n A_{\mathbf k,n}(0)$) at %the doping that corresponds to the van Hove singularity 
$p_{\text{vHs}}$ of each of the three band parameters.
The color in the legend specifies which curve and Fermi surface correspond to which band parameter. Yellow corresponds to the particle-hole symmetric case and black corresponds to an approximation to the band parameters of YBCO as calculated from density functional theory without interactions~\cite{pavarini_band-structure_2001,liechtenstein_quasiparticle_1996}. Hence, brown corresponds to an intermediate case, to show the effect of neglecting $t^{\prime\prime}$.
We define the reference $t=1$, which correspond to approximately 250 meV in YBCO. $p=1$ corresponds to an empty band (no electron), and $p=-1$ corresponds a completely full band. The triangles below the Fermi surface graphs show the doping corresponding to each Fermi surface, hence to each van Hove singularity. We see on the brown curve that this doping is not equal to the doping where $R_H=0$. The three dotted lines correspond to $p-1$, $p$ and $p+1$. }
\label{figureBare}
\end{figure}

Let us first look at the behavior of $n_H$ as a function $p$ for the bare band without any density wave
%$H=\sum_{\mathbf{k}}
%\epsilon_\mathbf{k} c^{\dag}_\mathbf{k} %c^{\vphantom{\dag}}_\mathbf{k}$.
($M=0$).
On Fig.~\ref{figureBare}, we present the Hall conductivity for three different band parameters $(t^\prime,t^{\prime\prime})$ and their corresponding Fermi surface at the van Hove singularity (vHs). This figure allows to understand three general facts.
%Introducing this figure let us understand three general facts. 

First, for any band parameter, a filled band (at $p\sim-1$) always behaves like a free hole gas 
%($n_H \sim p+1$) 
($n_H$ is the number of hole in the band) 
whereas an empty band (at $p\sim 1$) always behaves like a free electron gas 
%($n_H \sim -(p-1)$) 
($n_H$ is minus the number of electron in the band). Hence, $n_H$ change sign between $p=-1$ and $p=1$. It implies that at some doping $p_0$, the number of hole-like carriers must be equal to the number of electron-like carriers, hence $R_H=0$. When this happens, $n_H$ diverges. $R_H=0$ can happen for more than one doping as we will see in the next sections. Changing the band structure $\xi_\mathbf k$ only changes the value of the doping $p_0$, but the general behavior found in  Fig.~\ref{figureBare} is the same.

Second, although the doping $p_0$ %of zero Hall resistivity $p_{R_H}$
is always close to the doping of the van Hove singularity $p_{\text{vHs}}$, they are not always the same. The chemical potential corresponding to $p_{\text{vHs}}$ can be determined exactly by analytical calculation (Appendix~\ref{vanHoveSection}). %, and there is no limit in the numerical precision of the doping where $R_H=0$. 
Both dopings $p_{0}$ and $p_{\text{vHs}}$ are equal only for $(t^\prime,t^{\prime\prime})=(0,0)$. For $(t^\prime,t^{\prime\prime})=(-0.3,0.0)$, there is a clear offset between $p_{0}$ and $p_{\text{vHs}}$. However, the two dopings are always close because there are rapid changes in the Fermi surface near the van Hove singularity, causing rapid changes in the nature of charge carriers. 

%In fact, if we chose a scattering rate equal to the eigenstate velocity $\eta = \left| \nabla E_{\mathbf{k},n} \right|$ as in Refs.~\onlinecite{storey_hall_2016,verret_phenomenological_2017}, $p_{0}$ would be exactly $p_{\text{vHs}}$. This scattering rate would change the results presented, but we have good reasons to believe that a constant lifetime agrees better with Hall resistivity experiments (citations).

%However, the Fermi surface undergoes a reconstruction around the van Hove singularity. Hence 

Third, $t^{\prime\prime}$ has an impact as important as $t^{\prime}$ on the values of $p_{0}$ and $p_{\text{vHs}}$, so we must not neglect it. %(appendix~\ref{vanHoveSection}).

%Third, to correctly model the behavior of YBCO, we must not neglect $t^{\prime\prime}$ as we can see that it has an effect on the position of $R_H=0$ as important as $t^{\prime}$.

\subsection{Commensurate SDW ($\delta=0$)}

\begin{figure}[t]
	\includegraphics[width=1.00\columnwidth]{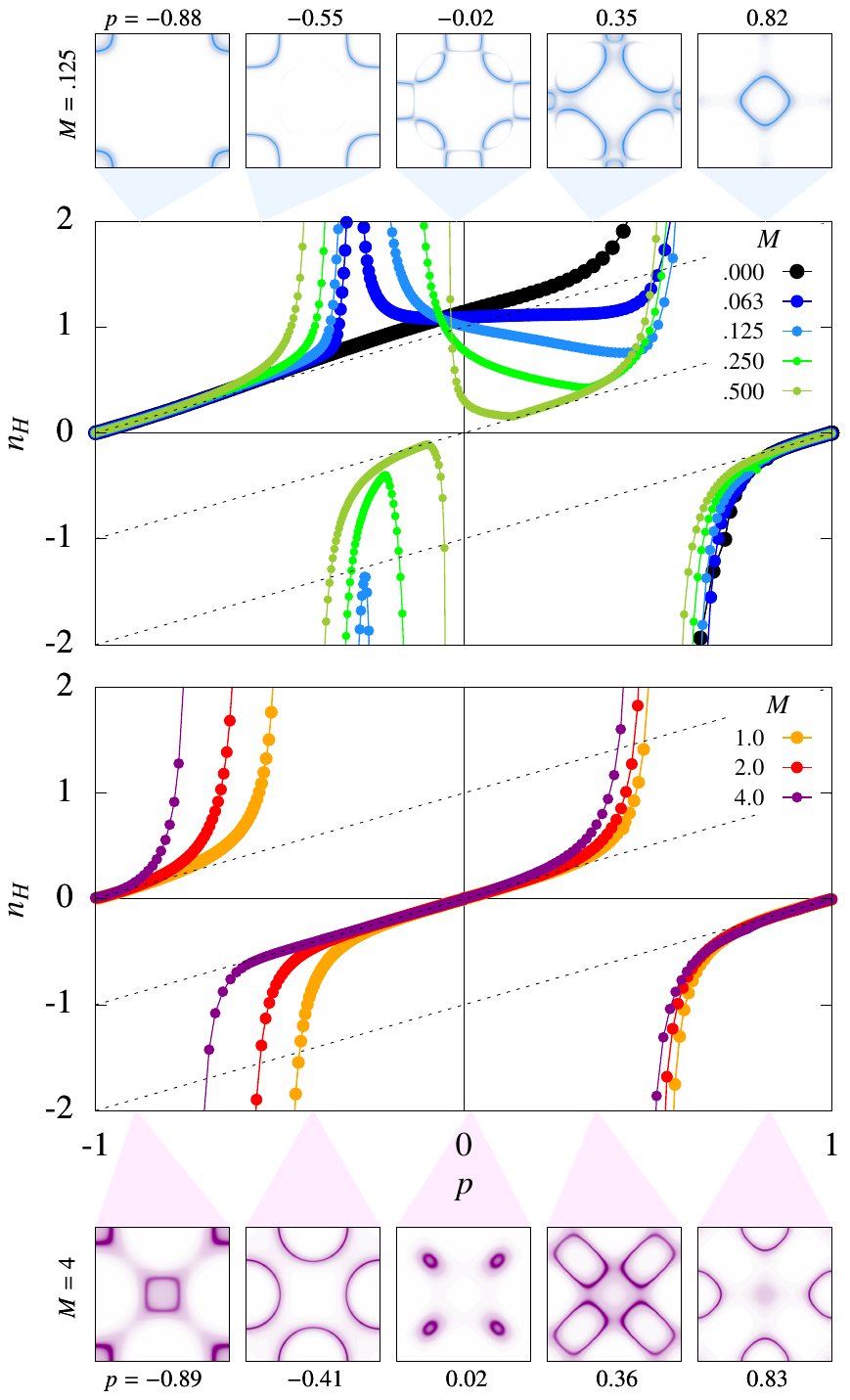}
	\caption{The middle two panels show the Hall number $n_H$ as a function of the hole doping $p$ relative to half-filling for different SDW amplitudes $M$ ($\delta=0$). The outer panels show the Fermi surface for five different dopings for $M=0.125$ (top blue Fermi surfaces) and for $M=4.0$ (bottom purple Fermi surfaces). The triangles below and above the Fermi surface graphs show the doping corresponding to each Fermi surface. The three dotted lines correspond to $p-1$, $p$ and $p+1$. Here, $t^{\prime}=-0.3$, $t^{\prime\prime}=0.2$. }
	\label{figureDelta0}
\end{figure}

Let us now look at the antiferromagnetic case. %defined by equations~\eqref{Hsdw} to~\eqref{Hsdw2}. 
In Fig.~\ref{figureDelta0}, we show the Hall number $n_H$ for different values of $M$ along with typical Fermi surfaces. % for small and big $M$ $M=0.0625$.  

%The black curve on both figure~\ref{figureBare} and~\ref{figureDelta0} is the same.

We separate the low and high SDW field $M$ on two different panels in order not to overload the plot. For low enough values of $M$, the $n_H$ curves deviate gradually from $M=0$ (black). The Fermi surface is almost equivalent to the bare Fermi surface, with additional anticrossing at the antiferromagnetic zone boundary.

For high field ($M\ge 1$), the system reaches another regime where $n_H$ is precisely proportional to $p$ near half-filling. This regime corresponds to an antiferromagnetic field $M$ so strong that it separates the original band in two new bands. Indeed, the curves are plotted as a function of the doping $p$, but if they were plotted as a function of the chemical potential, we would observe a gap at half-filling $p=0$; there would be a range of chemical potential with $n_H=0$. In fact, if we compare to the bare band behavior, we see that the pattern displayed in Fig.~\ref{figureBare} is repeated twice: hence the equivalence to two separated bands.

\begin{figure}[t]
	\includegraphics[width=1.00\columnwidth]{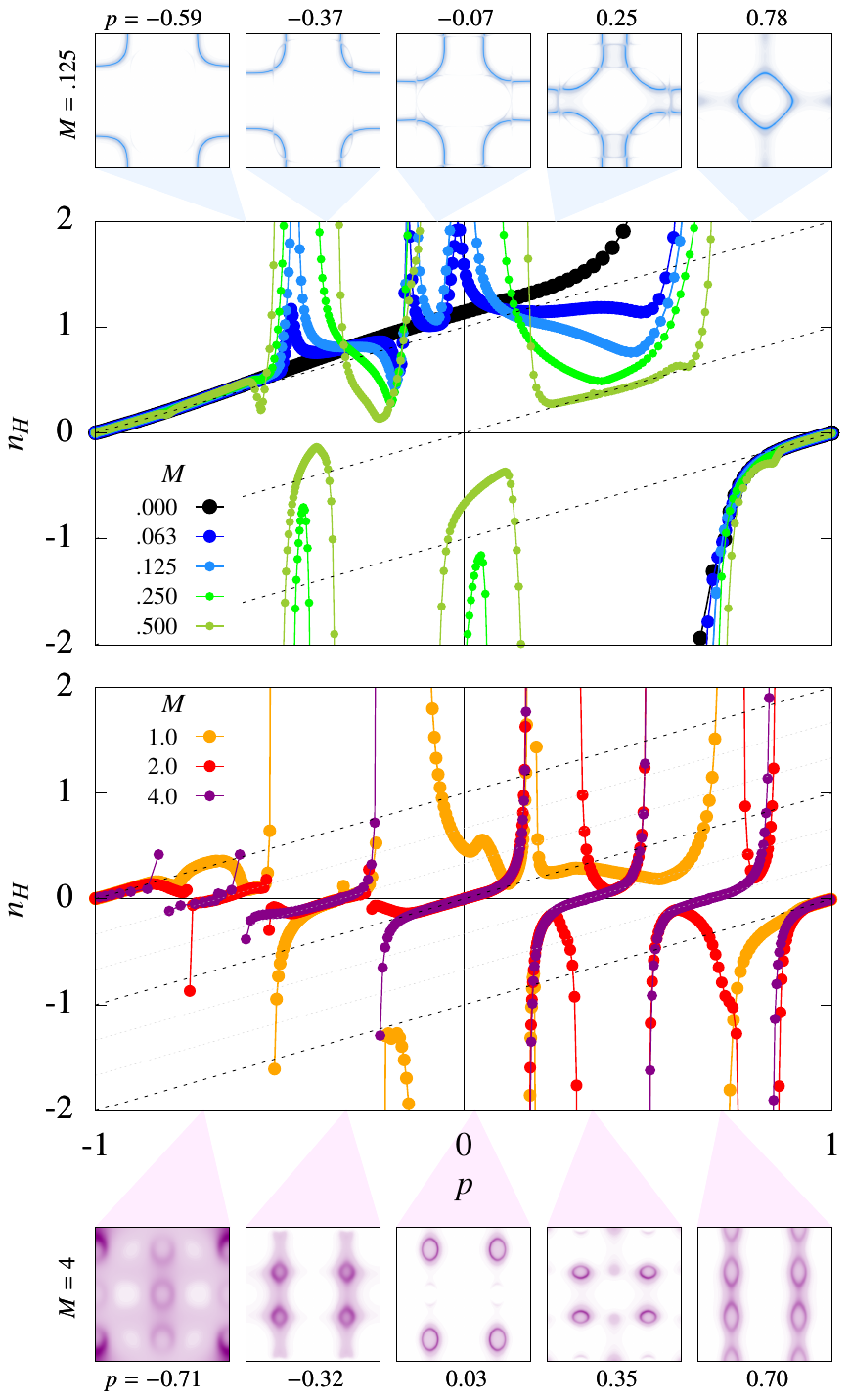}
	\caption{Same as Fig.~\ref{figureDelta0} but with $\delta=1/6$.}
	\label{figureDeltaFinite}
\end{figure}

From these results we can reproduce what was already published by Storey,~\cite{storey_hall_2016} simply by varying $M$ as a function of $p$. This corresponds to picking points from different curves depending on the value of $M(p)$ in the model. If we zoom on the portion $p=0$ to $p=0.3$ and vary the field $M$ linearly as a function of $p$, we also obtain that $n_H$ goes from $p$ to $1+p$. It is however not realistic to consider an antiferromagnetic regime for $p>0.05$ in hole-doped cuprates~\cite{hossain_situ_2008}. For this reason, we consider SDWs that are incommensurate and collinear, in the next section.

\subsection{Incommensurate SDW ($\delta=1/6$)}

On Fig.~\ref{figureDeltaFinite}, we show precisely the same quantities as in Fig.~\ref{figureDelta0}, but for incommensurate SDW with $\delta=1/6$.

For $M<1.0$, the behavior is similar to the antiferromagnetic case. For $M\ge 1.0$, the behavior is much more complicated but reaches a stable regime for $M\ge4.0$. We then recognize something similar to the $M=4.0$ curve of Fig.~\ref{figureDelta0}: the strong field $M$ causes multiple band splittings. %caused by the strong field $M$. 
In fact, we have precisely $2L=6$ bands when $\delta=1/6$. Indeed, the original band (for $M=0$) in the original Brillouin zone is split into $6$ different portions, one for each rBz. Every rBz contains the same number of k-points, hence the same number of states. In fact, $n_H$ vanishes precisely when $p$ is a multiple of $1/3$, in other words, $1/6$ of the total band electron. As $M$ increases, these $6$ bands separate completely from each other. The Hall number $n_H$ thus presents signatures of these $6$ independent bands. Note that if we chose $\delta=1/8$ we would have $8$ different bands, and so on.

From this observation, we can conclude that for the incommensurate case, the regime $M \gg t$ is unlikely to be the cause of the $p$ behavior near half-filling, contrary to the commensurate case. Indeed, even if we find that $n_H \sim p$ on Fig.~\ref{figureDeltaFinite} around half-filling, the region over which we find this behavior decreases with $2L_y$. As the fractions $\delta$ considered are more and more incommensurate, in other words as $2L_y$ increases, $n_H$ is more and more constrained to zero when $M$ is large. Thus we lose this $n_H \sim p$ behavior for incommensurate SDW at large~$M$. % since $\delta$ varies as a function of doping.

One could argue that, even if the $n_H \sim p$ behavior around half-filling is not obtained at high $M$, it could somehow appear at intermediate $M$, like the curve $M \sim 0.5$ of Fig.~\ref{figureDeltaFinite} seems to suggest. We study this case in the next section.

\subsection{Constant $M$, different $\delta$ }

\begin{figure}[t]
\includegraphics[width=1.00\columnwidth]{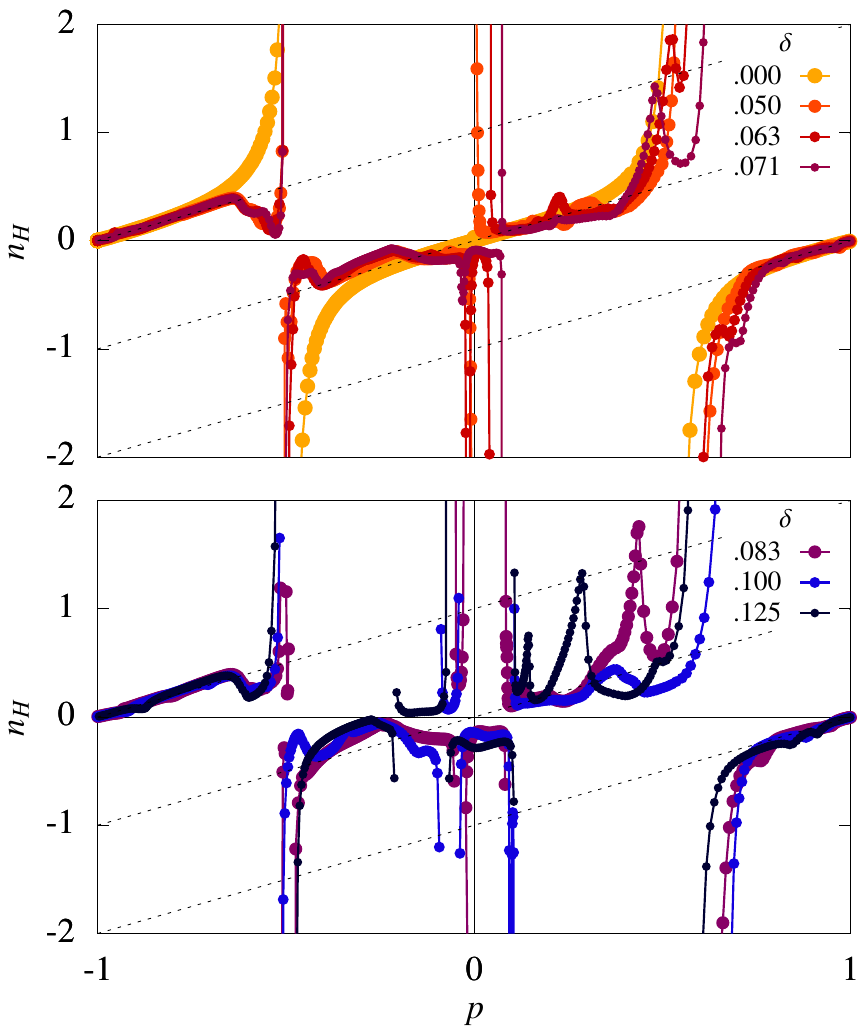}
\caption{Hall number $n_H$ as a function of the hole doping $p$ relative to half-filling for different SDW incommensurability $\delta$ with the same $M=1.0$. We separate small and big $\delta$ on two panels to lighten the plot. The $\delta$ used are $0$, $\frac{1}{20}=0.050$, $\frac{1}{16}=0.063$, $\frac{1}{14}=0.071$, $\frac{1}{12}=0.083$, $\frac{1}{10}=0.100$, $\frac{1}{8}=0.125$. $t^{\prime}=-0.3$, $t^{\prime\prime}=0.2$.}
\label{figureM1}
\end{figure}

Fig.~\ref{figureM1} shows how the parameter $\delta$ influences $n_H$. We choose $M=1.0$ because it is not too large but still sufficient to find the $n_H \sim p$ behavior around half-filling for the commensurate antiferromagnetic case ($\delta=0$) (see Fig.~\ref{figureDelta0}). On Fig.~\ref{figureM1}, we see that, for small $\delta$, the result is close to the commensurate case. In other words, we find a behavior close to $n_H \sim p$ near half-filling. However, the more we increase $\delta$, the more it deviates from $n_H \sim p$. 

In the experiments, $\delta$ is a function of $p$~\cite{yamada_doping_1998,haug_neutron_2010,fujita_static_2002}. Therefore, a natural question arises: is it possible to observe $n_H \sim p$ at low doping if we vary $\delta$ as a function of $p$?

% If we vary $\delta$ as a function of $p$, as observed in the experiments~\cite{yamada_doping_1998,haug_neutron_2010,fujita_static_2002}, will it be possible to observe $n_H \sim p$ at low doping? 

\subsection{Variable $\delta$ as a function of $p$ }

\begin{figure}[t]
\includegraphics[width=1.00\columnwidth]{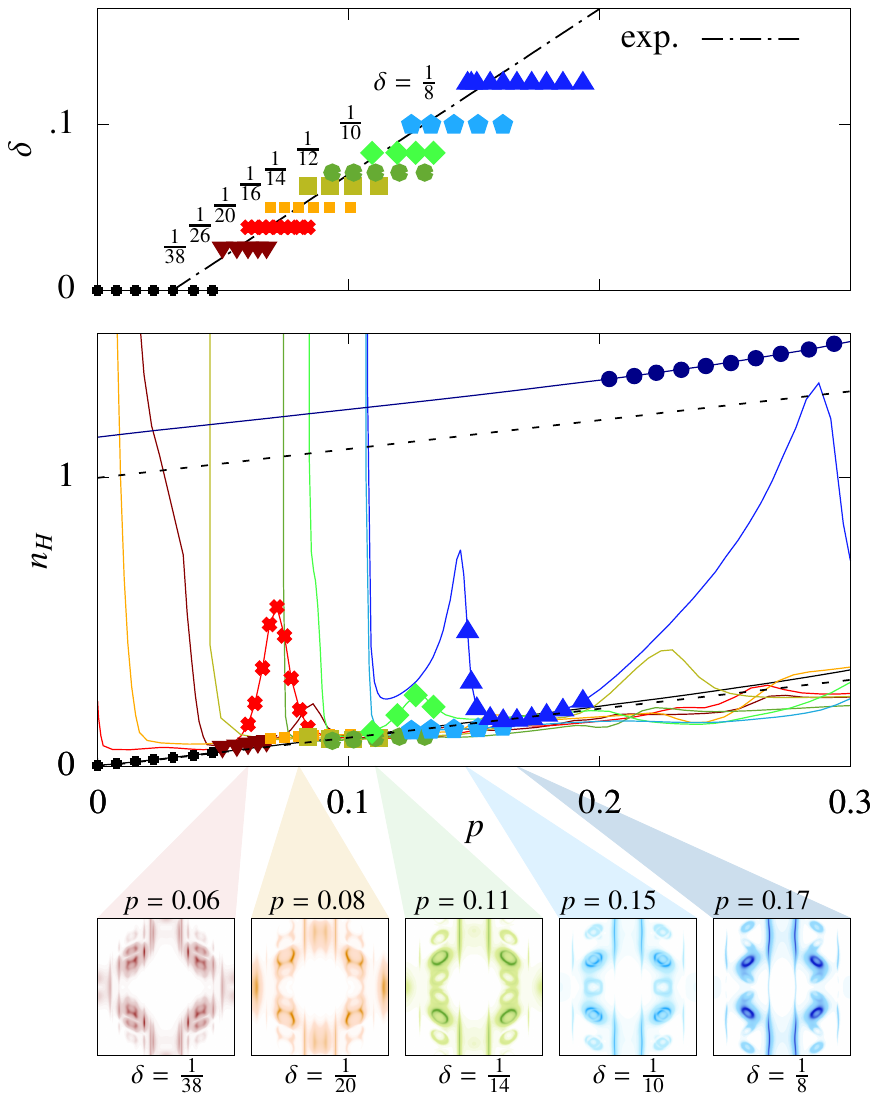}
\caption{The top panel shows the chosen distribution of $\delta$ values as function of $p$. Each symbol of a given color corresponds to the same incommensurability $\delta$: $0$, $\frac{1}{38}=0.026$, $\frac{1}{26}=0.038$, $\frac{1}{20}=0.050$, $\frac{1}{16}=0.063$, $\frac{1}{14}=0.071$, $\frac{1}{12}=0.083$, $\frac{1}{10}=0.100$ and $\frac{1}{8}=0.125$. The dotted line corresponds to the experimental reference $\delta \sim p-0.03$.~\cite{haug_neutron_2010} In the middle panel, Hall number $n_H$ as a function of hole doping $p$ relative to half-filling for different SDW incommensurability $\delta$. We use $M=1.0$ below $p=0.2$ and $M=0.0$ above. Each curve corresponds to a different $\delta$. The symbols on the middle panel correspond to the symbols on the top panel. We only draw symbols on the portion of the curve that corresponds to the experimental reference $\delta \sim p-0.03$. We also show the corresponding curve for the whole $p$ range for each $\delta$ in order to highlight the deviations from $n_H\sim p$ close to the data points selected. In the bottom panels we show Fermi surfaces for five different $p$ and $\delta$. The triangles above the Fermi surface graphs show the doping corresponding to each Fermi surface.}
\label{figureDeltaVariable}
\end{figure}

%In experiments, the incommensurability $\delta$ changes as a function of doping $p$~\cite{yamada_doping_1998,haug_neutron_2010,fujita_static_2002}. 
%%What would the $n_H$ curve look like if we reproduce the correct $\delta(p)$ behavior? The result is presented on
In Figs.~\ref{figureDeltaVariable}, we look at the $n_H$ curve when the experimentally observed $\delta(p)$ is used. We see that for these choices of $\delta$ as a function of $p$, $n_H$ has a tendency to follow $p$, but is not locked to $p$. Note that we chose $\delta \sim p -0.03$ as reported by experiments on YBCO \cite{haug_neutron_2010}, but choosing a slightly different $\delta$ dependency leads to the same conclusion. Also, as discussed before, increasing or decreasing $M$ would make $n_H$ deviate even more from the $p$ behavior around half-filling. We claim here that, for the model used, the parameters used in Figs.~\ref{figureDeltaVariable} are close to the best set of parameters to reproduce the $n_H \sim p$ behavior, yet it still lacks agreements with experiments~\cite{badoux_change_2016,segawa_intrinsic_2004,ando_evolution_2004}. % However, earlier experiments on LSCO thin films show similar structures~\cite{balakirev_quantum_2009}.

On the same figure, we see that the Fermi surfaces corresponding to the best-case scenario are different from the ARPES results on YBCO~\cite{hossain_situ_2008,meng_coexistence_2009}. %We see four pockets % with stripes 
%and only one anti-node region that is not gapped. These Fermi surfaces are highly anisotropic. 
In the experiments, if we ignore the effects of bilayer splitting and copper oxide chains, the Fermi surface only consists of four Fermi arcs. It is often speculated that those arcs are in fact four small pockets with their back spectral weight too faint to be measured. By contrast, in our computed Fermi surfaces, spectral weight remains in the $y$ antinodes, and many copies of the nodal pocket appear along the $y$ direction. These features are absent in experiments. %This is closely related to our choice of a unidirectional SDW (Eq.~\eqref{unidirectionalQ}). 

\section{Discussion}
\label{discussion}

Our analysis indicates that the incommensurate collinear SDW, as represented by our model, cannot explain the $n_H \sim p$ behavior of the underdoped YBCO measurements~\cite{badoux_change_2016,segawa_intrinsic_2004}. Even the best-case scenario ($M=1$ and  $\delta \sim p-0.03$, as in Fig.~\ref{figureDeltaVariable}) %does not reproduce exactly the $n_H \sim p$ behavior. It 
predicts important deviations from the $n_H \sim p$ behavior. %If incommensurate collinear SDW was responsible for the $n_H \sim p$ behavior, we would be able to observe these deviations in YBCO Hall measurements~\cite{badoux_change_2016,segawa_intrinsic_2004}. 
Those deviations were not seen in experiments on YBCO~\cite{badoux_change_2016,segawa_intrinsic_2004}.
However, Hall measurements in LSCO and BLSCO~\cite{balakirev_quantum_2009} report a sharp feature in $n_H$ around $p=0.16$ reminiscent of the deviation from $n_H\sim p$ we predict here, but the authors argue this feature is linked to the high temperature superconducting mechanism. It could be worth investigating if this feature is not rather linked to density waves similar to those studied here.

%Even though this conclusion seems definitive, there is still a caveat. 

%Our calculation considered long range SDW even if experiments indicate a short range SDW for $p > 0.05$. %Even if we choose a $\delta$ with a very large denominator (thus a large $L_y$), the pattern will repeat every $L_y$ unit cell, since we need to use periodic conditions to calculate the spectrum of the Hamiltonian. 
%We do not expect it would change a lot the conclusions, but one must be aware of this detail.

Note also that we chose the values of $M$ that provided the desired qualitative behavior of $n_H \sim p$. It does not imply any quantitative prediction. What we called the ``best case scenario'' (Figs~\ref{figureDeltaVariable}) is not a proof that $M$ should have a value around~$1$, which is $t$ in our units. In fact, the $M$ corresponding to the real SDW found in YBCO should be much smaller than $M\sim1$, since for such large values of $M$, there are strong irregularities in the calculated Fermi surfaces, as shown on Fig.~\ref{figureDeltaVariable}.

The results for the commensurate antiferromagnetic case shown on Fig.~\ref{figureDelta0} can, however, explain the Hall effect measurements in electron-doped Pr$_{2−x}$Ce$_x$CuO$_{4}$~\cite{dagan_fermi_2016} and  La$_{2−x}$Ce$_x$CuO$_{4}$~\cite{sarkar_fermi_2017}, where $x$, the doping in electrons, corresponds to $x=-p$ on the electron doped side. Indeed, starting from $p=0$, if we decrease $M$ as we decrease $p$ (increase $x$), at some point  there will be a sign change in $n_H$, as observed experimentally in Refs.~\cite{dagan_fermi_2016,sarkar_fermi_2017}. This is one possible explanation of the $p$ to $1+p$ (or $-x$ to $1-x$) transition in the electron-doped cuprates. 
Note however, as argued in Refs.~\cite{dagan_fermi_2016,sarkar_fermi_2017}, that the proximity of the vHs to $p=0$ remains a plausible alternative explanation.

We must stress that the conclusions reached here, with the incommensurate collinear SDW, do not extend to the incommensurate spiral SDW, as both orders are fairly different (they are only the same for $\mathbf Q=(\pi,\pi))$. Indeed, this explains the significant difference between the results of our model and the results of Eberlein et al.~\cite{eberlein_fermi_2016-1}. 
The local moment in a spiral SDW is constant in magnitude but its direction rotates, whereas the local moment in a collinear SDW has a fixed direction but its amplitude is modulated. It is also possible to model a truly incommensurate $\mathbf Q$ spiral with a $2\times2$ matrix (two-band model), without any approximation~\cite{eberlein_fermi_2016-1}.  The value of $\mathbf Q$ can be as incommensurate as needed. With the method presented in this article, a truly incommensurate collinear SDW would necessitate a matrix of infinite size. 

A natural extension of this study would be to average over a spread in $\mathbf Q$ vector to simulate shorter range correlations. %Indeed, the peaks in the $n_H$ curve are associated with the fine details of the Fermi surface . 
Indeed, in the large $M$ limit, as shown on Fig.~\ref{figureDeltaFinite}, there are precisely $2L+1$ peaks (at values of $p$ where $R_H=0$), which is an artifact of commensurability in our approach. So averaging over a spread in $\mathbf Q$ vector would smear out the fine details of the Fermi surface and possibly smear the peaks in $n_H$ (Fig.~\ref{figureDeltaVariable}), resulting in $n_H \sim p$ behavior. Adding disorder to the model would probably result in a similar effect. This is outside the scope of the model presented here, but it would be interesting to verify this point in future work. In any case, there are multiple refinements needed to reproduce the $n_H\sim p$ behavior with incommensurate collinear SDW, whereas two-band effective models~\cite{storey_hall_2016,eberlein_fermi_2016-1,chatterjee_fractionalized_2016,verret_phenomenological_2017} do not need any impurities, $\mathbf Q$ averaging, or fine-tuned value of the effective mean field $M$ to obtain the $n_H\sim p$. 
%The conclusion of our analysis goes beyond the incommensurate collinear SDW.
From previous studies of these models
~\cite{storey_hall_2016,eberlein_fermi_2016-1,chatterjee_fractionalized_2016,verret_phenomenological_2017}, we know that two-band effective models are sufficient to obtain the $n_H \sim p$ behavior because they open a gap at half-filling. By contrast, the incommensurate SDW studied here splits the dispersion in more than two bands, which causes deviations from the seeked $n_H \sim p$ behavior. %The conclusion of our analysis goes beyond the incommensurate collinear SDW. 
We can infer that the opening of a gap at half-filling might be an important necessary feature of any adequate theory of the zero-temperature normal state in the pseudogap regime. %This last statement goes beyond the incommensurate collinear SDW. 
Nonetheless, the actual physics behind the pseudogap at zero temperature is probably more subtle, being deeply rooted in strongly correlated physics as indicated by methods like cluster perturbation theory~\cite{senechal_cluster_2002,senechal_hot_2004} or generalizations of dynamical mean-field theory to clusters, like cellular dynamical mean-field theory CDMFT or the dynamical cluster approximation DCA~\cite{kotliar_cellular_2001,lichtenstein_antiferromagnetism_2000,civelli_dynamical_2005,kyung_pseudogap_2006,stanescu_fermi_2006,macridin_pseudogap_2006,haule_strongly_2007,kancharla_anomalous_2008,ferrero_pseudogap_2009,ferrero_valence_2009,sakai_evolution_2009,gull_momentum-space_2010,sordi_mott_2011}. %The pseudogap found with these methods is sometimes presented as Mott physics that gap only a part of the $\mathbf k$-space (the antinodes). With the analysis presented here, what the $n_H \sim p$ behavior seen in experiments may be revealing is that precisely half the Brillouin zone is lost when entering the pseudogap state, no more, no less. This statement obviously requires more investigation.
It would be interesting to calculate the value of $n_H$ for the pseudogap regime with these techniques in future work. %Care must be taken to included vertex corrections 

\section{Acknowledgments}
\label{acknowledgments}

We acknowledge S. Badoux, I. Garate, R. Nourafkan and L. Taillefer for discussions. This work was partially supported
by the Natural Sciences and Engineering Research Council
(Canada) under Grant No. RGPIN-2014-04584 and the Research Chair on the Theory of Quantum Materials (A.-M.S.T.). Simulations were
performed on computers provided by the Canadian Foundation
for Innovation, the Ministere de l'\'Education des Loisirs et du
Sport (Qu\'ebec), Calcul Qu\'ebec, and Compute Canada.

% % % % % % % % % % % % % % % % % % % % % % % % %
% % % % % % % % % % % % % % % % % % % % % % % % %
% % % % % % % % % % % % % % % % % % % % % % % % %
% % % % % % % % % % % % % % % % % % % % % % % % %
% % % % % % % % % % % % % % % % % % % % % % % % %
% % % % % % % % % % % % % % % % % % % % % % % % %

\appendix
\section{Derivatives of eigenenergies}
\label{derivativesSection}
Expressions for %In order to calculate 
the conductivities $\sigma_{xx}$ and $\sigma_{xy}$ contain %, we need to determine 
the derivative of the eigenenergies of the Hamiltonian: $\tfrac{\partial E_{\mathbf{k},n}}{\partial k_\alpha}$ and $\tfrac{\partial^2 E_{\mathbf{k},n}}{\partial k_\alpha\partial k_\beta}$. For a two-band model, like the antiferromagnet, the calculation is straightforward. However, for larger matrices (size 3 or more), the analytic expression of the eigenvalues is much more complicated and one must rely on a numerical approach. We could find the derivatives with finite differences but there is imprecision around degeneracies due to the arbitrary ordering of the eigenenergies $E_{\mathbf{k},n}$ (for some specific $\mathbf{k}$ points). It is important to optimize this diagonalization since it is the bottleneck of the calculation for large $L_y$. In this appendix, we present a general and straightforward approach to calculate exactly these derivatives for any $\mathbf{k}$.

\subsection{First derivative}
The first derivative $\tfrac{\partial E_{\mathbf{k},n}}{\partial k_\alpha}$ is obtained from the Hellmann-Feynman theorem. Here we recall the proof.
%easy to calculate. 
Starting from the eigenequation (ignoring the spin index here):
\begin{align}
\hat{H}_\mathbf{k} |\psi_{\mathbf{k},n}\rangle = E_{\mathbf{k},n} |\psi_{\mathbf{k},n}\rangle,
\label{valeur_propres}
\end{align}
where $\mathbf{k}$ is the wavevector, $\hat{H}_\mathbf{k}$ is the Hamiltonian operator, $E_{\mathbf{k},n}$ are the eigenenergies corresponding to the eigenstates $|\psi_{\mathbf{k},n} \rangle$. Let us drop the explicit $\mathbf{k}$ in the notation from here. The eigenbasis is orthonormal:
\begin{align}
\langle \psi_n | \psi_m \rangle &= \delta_{n,m}
\\
\frac{\partial}{\partial k_{\alpha}} \langle \psi_n | \psi_n \rangle
&=
\frac{\partial \langle \psi_n |}{\partial k_{\alpha}}  | \psi_n \rangle 
+ \langle \psi_n| \frac{\partial | \psi_n \rangle}{\partial k_{\alpha}}
= 0.
\label{derivee_ortho}
\end{align} 
%Let us take the derivative of $\langle \psi_n|$ times \eqref{valeur_propres}:
Multiplying~\eqref{valeur_propres} by $\langle \psi_n|$ and taking the derivative, we obtain: %times
\begin{align}
\frac{\partial}{\partial k_\alpha} \langle \psi_n | \hat{H} | \psi_n \rangle
=&\hphantom{+} 
\langle \psi_n | \frac{\partial \hat{H}}{\partial k_\alpha} | \psi_n \rangle
\label{hellmann_feynman0}
\\&+
\underbrace{
\frac{\partial \langle \psi_n | }{\partial k_\alpha} E_n | \psi_n \rangle
+\langle \psi_n | E_n \frac{\partial  | \psi_n \rangle}{\partial k_\alpha}
}_{=0}.
\nonumber
\end{align} 
The last two terms vanish %of equation~\eqref{hellmann_feynman0} are zero 
because of Eq.~\eqref{derivee_ortho}. Using the eigenequation on the left-hand term, we find:
\begin{align}
\frac{\partial E_n}{\partial k_\alpha}
&= 
\langle \psi_n | \frac{\partial \hat{H}}{\partial k_\alpha} | \psi_n \rangle
\label{hellmann_feynman}
\end{align}
which is known as the Hellmann-Feynman theorem~\cite{feynman_forces_1939,deb_note_1972}. 

\subsection{Second derivative}

For the second derivative, the approach is similar. %Let us take the 
Taking the derivative of equation~\eqref{hellmann_feynman}, we obtain three terms:
\begin{align}
\frac{\partial^2 E_n}{\partial k_\beta \partial k_\alpha}
=&\hphantom{+}
\langle \psi_n | 
\frac{\partial^2 \hat{H}}{\partial k_\beta \partial k_\alpha} 
| \psi_n \rangle
\nonumber
\\&+
\frac{\partial \langle \psi_n | }{\partial k_\beta}
\frac{\partial \hat{H}}{\partial k_\alpha}
| \psi_n \rangle
+
\langle \psi_n |
\frac{\partial \hat{H}}{\partial k_\alpha}
\frac{\partial  | \psi_n \rangle}{\partial k_\beta}.
\label{hellman_feynman_checkpoint1}
\end{align}
The first term is straightforward to calculate but the last two terms must be further simplified. The derivative with respect to $k_\alpha$ of the eigenequation~\eqref{valeur_propres} can be reordered as: %and
%\begin{align}
%\frac{\partial}{\partial k_\alpha} 
%\bigg( 
%\hat{H} |\psi_n\rangle 
%\bigg)
%&= 
%\frac{\partial}{\partial k_\alpha} 
%\bigg(
%E_n |\psi_n\rangle 
%\bigg)
%\end{align}
%reordering the terms, we get:
\begin{align}
\frac{\partial\hat{H}}{\partial k_\alpha} 
|\psi_n\rangle 
&=
\frac{\partial E_n}{\partial k_\alpha} 
|\psi_n\rangle
-
\big( \hat{H} - E_n \big)
\frac{\partial|\psi_n\rangle}{\partial k_\alpha}.
\label{hellman_feynman_checkpoint2}
\end{align}
which can be substituted twice in equation~\eqref{hellman_feynman_checkpoint1}. %After the subsitution of~\eqref{hellman_feynman_checkpoint2} in~\eqref{hellman_feynman_checkpoint1}, 
We obtain multiple terms, two of which cancel due to equation~\eqref{derivee_ortho}: % In the end, we get:
\begin{align}
\frac{\partial^2 E_n}{\partial k_\beta \partial k_\alpha}
&= 
\langle \psi_n | 
\frac{\partial^2 \hat{H}}{\partial k_\beta \partial k_\alpha} 
| \psi_n \rangle
-
2 \frac{\partial \langle \psi_n | }{\partial k_\beta}
\big( \hat{H} - E_n \big)
\frac{\partial|\psi_n\rangle}{\partial k_\alpha} .
\label{hellmann_feynman_checkpoint3}
\end{align}
Note that we could not isolate $\frac{\partial|\psi_n\rangle}{\partial k_\alpha}$ directly in Eq.~\eqref{hellman_feynman_checkpoint2} because by definition of the eigenvalues $E_n$, the determinant of $( \hat{H} - E_n )$ is zero, thus $( \hat{H} - E_n )$ cannot be inverted.
%This 

Form~\eqref{hellmann_feynman_checkpoint3} is simpler, but we still need to determine correctly the derivative of the eigenstate $\frac{\partial|\psi_n\rangle}{\partial k_\alpha}$. This can be calculated exactly using perturbation theory. Starting with the definition of the derivative (in one dimension $k$ for simplicity):
\begin{align}
\frac{\partial|\psi_n (k)\rangle}{\partial k} 
&=
\lim\limits_{\delta k \rightarrow 0}
\frac{
|\psi_n (k + \delta k)\rangle - |\psi_n (k)\rangle 
}{ \delta k }.
\label{def_derive}
\end{align}

%Let us consider 
The Hamiltonian $\hat{H}(k + \delta k)$ has $|\psi_n (k + \delta k)\rangle$ as eigenstates. Since $\delta k$ is small by definition and the eigenenergies vary smoothly as a function of $k$, the Hamiltonian at $k + \delta k$ can be expressed by a small perturbation from $\hat{H}(k)$:
\begin{align}
\hat{H}(k+\delta k)
=
\hat{H}(k) + 
\underbrace{
[\hat{H}(k+\delta k)-\hat{H}(k)]
}_{\text{perturbation}}\,.
\end{align}
%By definition of the derivative (equation~\eqref{def_derive}), we know that 
Since $\delta k \rightarrow 0$ in Eq.~\eqref{def_derive}, the first order perturbation term of %hence the linear perturbation definition 
$|\psi_n (k + \delta k)\rangle$ is exact:
\begin{align}
&|\psi_n (k + \delta k)\rangle 
-|\psi_n (k)\rangle
= 
\label{checkpoint0}
\\
\nonumber 
&\sum_{m\neq n}\frac{
\langle\psi_m (k)|
\Big(\hat{H}(k + \delta k)-\hat{H}(k)\Big)
|\psi_n (k)\rangle
}{E_n(k) - E_m(k)}
|\psi_m (k)\rangle .
\end{align}
Substituting equation~\eqref{checkpoint0} into \eqref{def_derive}, we obtain:
%\begin{align}
%\frac{\partial|\psi_n (\mathbf{k})\rangle}{\partial k_\alpha} 
%=
%\sum_{m\neq n}\frac{
%\langle\psi_m (\mathbf{k})|
%\bigg(
%\lim\limits_{\delta k_\alpha \rightarrow 0} 
%\frac{
%\hat{H}(\mathbf{k}+\delta k)-\hat{H}(\mathbf{k})
%}{ \delta k } \bigg)
%|\psi_n (\mathbf{k})\rangle
%}{E_n(\mathbf{k}) - E_m(\mathbf{k})}
%|\psi_m (\mathbf{k})\rangle
%\end{align}
%Cleaning up and omitting the explicit dependency over $\mathbf{k}$, we get:
\begin{align}
\frac{\partial|\psi_n\rangle}{\partial k_\alpha} 
&=
\sum_{m\neq n}\frac{
\langle\psi_m|
\frac{\partial\hat{H}}{ \partial k_\alpha }
|\psi_n\rangle
}{E_n - E_m}
|\psi_m\rangle
\label{derivative_eigenvector}
\end{align}
%Equation~\eqref{derivative_eigenvector} 
which holds for any dimension of $k$ space. 
%Combined with equation~\eqref{hellmann_feynman_checkpoint3}, we can calculate exactly the second derivative of the eigenenergies for any $\mathbf{k}$ point. 
This formula is commonly used in the calculation of the Berry connection~\cite{xiao_berry_2010,berry_quantal_1984}. 

Substituting in Eq.~\eqref{hellmann_feynman_checkpoint3}, we find the band version of the {\it f-}sum rule \cite{wilson_1953,Fukuyama:1971}:
\begin{align}
\frac{\partial^2 E_n}{\partial k_\beta \partial k_\alpha}
=&\hphantom{+}
\langle \psi_n | 
\frac{\partial^2 \hat{H}}{\partial k_\beta \partial k_\alpha} 
| \psi_n \rangle \nonumber
\\
&+2
\sum_{m\neq n}\frac{
\langle\psi_n|
\frac{\partial\hat{H}}{ \partial k_\beta }
|\psi_m\rangle
\langle\psi_m|
\frac{\partial\hat{H}}{ \partial k_\alpha }
|\psi_n\rangle
}{E_n - E_m}\,.
\end{align}

%\subsection{Matrix notation}
%This derivation is useful since the corresponding expression for a Hamiltonian $\hat{H}_\mathbf{k}$ represented in matrix notation is straightforward for any matrix dimension. Let $\hat{U}_\mathbf{k}$ be the unitary matrix that diagonalizes $\hat{H}_\mathbf{k}$. Omitting the label $\mathbf{k}$, the Hellmann-Feynman theorem, Eq.~\eqref{hellmann_feynman}, can be expressed as
%\begin{align}
%\frac{\partial E_n}{\partial k_\alpha} 
%&=
%\Big( 
%\hat{U}^{\dag}
%\frac{\partial \hat{H}}{ \partial k_\alpha }
%\hat{U}^{\vphantom{\dag}}
%\Big)_{nn},
%\label{derivative_eigenvector_matrix1}
%\end{align}
%and the second derivative as
%\begin{align}
%\frac{\partial^2 E_n}{\partial k_\beta \partial k_\alpha}
%&= 
%\Big(
%\hat{U}^{\dag}
%\frac{\partial^2 \hat{H}}{\partial k_\beta \partial k_\alpha} 
%\hat{U}^{\vphantom{\dag}}
%-
%2 \frac{\partial \hat{U}^{\dag} }{\partial k_\beta}
%\big( \hat{H} - E_n \big)
%\frac{\partial \hat{U}^{\vphantom{\dag}} }
%{\partial k_\alpha} 
%\Big)_{nn},
%\label{derivative_eigenvector_matrix2}
%\end{align}
%with
%\begin{align}
%\frac{\partial U_{ln}}{\partial k_\alpha} 
%&=
%\sum_{\substack{m\neq n\\ij}}
%U_{lm}
%\frac{
%U^{*}_{im}
%\frac{\partial H_{ij}}{ \partial k_\alpha }
%U^{\vphantom{*}}_{jn}
%}{E_n - E_m}.
%\label{derivative_eigenvector_matrix3}
%\end{align}
Since the derivatives of the Hamiltonian matrix are easy to obtain analytically, the only numerical part of the calculation which must be performed at any $\mathbf{k}$ point is the diagonalization of the Hamiltonian to obtain %the unitary matrix and 
the eigenvalues. %is required.

\section{van Hove singularity energy}
\label{vanHoveSection}
Here, we derive a simple equation to find the energy corresponding to the van Hove singularity in a tight-binding model with the dispersion in Eq.~\eqref{basicDispersion0}.

The van Hove singularities occur at $\frac{ \partial \xi_{\mathbf{k}}}{ \partial k_x} = \frac{ \partial \xi_{\mathbf{k}}}{ \partial k_y} = 0$, where:
\begin{align}
\frac{ \partial \xi_{\mathbf{k}}}{ \partial k_x} = 
\hphantom{+} &2t \sin (k_x) +4t^{\prime\prime} \sin (2 k_x) \nonumber \\
+ &2t^{\prime} (\sin (k_x + k_y) + \sin (k_x - k_y)) 
\end{align}
and similar for $\frac{ \partial \xi_{\mathbf{k}}}{ \partial k_y}$.  If we focus on the singularities that can be found on the axis $k_y=0$, we obtain that the $k_x$ where the saddle point in the energy is:
\begin{align}
\tilde{k}_x &= \left\lbrace    {\begin{array}{ll}
   \arccos (r) &\mathrm{~~~if~} |r| \leq 1  \\
   \pi & \mathrm{~~~otherwise}  
\end{array} }\right.
\label{kxTilde}
\\
\text{where}\;\;\; r &\equiv -\frac{t+t^{\prime}}{4 t^{\prime\prime}}.
\end{align}
Using some trigonometric identities on equation~\eqref{basicDispersion0} together with equation~\eqref{kxTilde}, we can calculate that the energy corresponding to this saddle point is
\begin{align}
\xi_{\mathbf{k}=(\tilde{k}_x,0)} = 
 \left\lbrace    {\begin{array}{ll}
   -2t(1+r)-4t^{\prime}r -4t^{\prime\prime}r^2 &\mathrm{~~~if~} |r| \leq 1  \\
   -4t^{\prime} -4t^{\prime\prime}& \mathrm{~~~otherwise}  
\end{array} }\right. .
\label{dispersionVanHove}
\end{align}
This reduces to the result of Ref.~\cite{benard_magnetic_1993} when $t^{\prime\prime}=0$. Note that to be general, we would need to search for singularities that cannot be found on the axis $k_x=0$ or $k_y=0$, but when $t^{\prime}$ and $t^{\prime\prime}$ are small compared to $t$, as in every cuprate material, the saddle points can only be found on the axis $k_x=0$ or $k_y=0$. This can be seen on the Fermi surfaces of Figs~\ref{figureBare}: the van Hove singularities 
%in other words, the cross in the Fermi surfaces 
are only found on the axis $k_x=0$ and $k_y=0$.

\end{document}